\documentclass{ar-1col-S2O}
\usepackage[numbers]{natbib}
\usepackage{url}
\usepackage{amsmath}
\usepackage{amssymb}
\newcommand{\beq} {\begin{equation}}
\newcommand{\eeq} {\end{equation}}
\newcommand{\bea} {\begin{eqnarray}}
\newcommand{\eea} {\end{eqnarray}}
\newcommand{\be} {\begin{equation}}
\newcommand{\ee} {\end{equation}}

\jname{Annual Review of Condensed Matter Physics}
\jvol{18}
\jyear{2026}
\doi{10.1146/((please add article doi))}

\begin{document}

\markboth{Abanov $\bullet$ Chubukov}{Non-BCS Superconductivity}

\title{Non-BCS Pairing by a Singular Dynamical Interaction}

\author{ Artem G. Abanov$^1$ and Andrey V. Chubukov$^2$
\affil{$^1$Department of Physics and Astronomy, Texas A\&M University, College Station, TX 77843;
email: abanov@tamu.edu}
\affil{$^2$School of Physics and Astronomy, University of Minnesota, Minneapolis, MN 55455;
email: achubuko@umn.edu}}

\begin{abstract}
This review examines the theory of superconductivity in systems with {\em singular dynamical} electron-electron interaction
and contrasts  it with a conventional  BCS superconductivity. Examples include metals near a Quantum Critical Point, quantum dots and system near a localization (Mott) transition. We show, that  the  singular interaction destroys the traditional separation of energy scales, invalidating the significance of Cooper logarithm, and, as the consequence, the whole  
 BCS framework. 
 We explore the universal model with dynamical interaction $\Gamma (\Omega) \propto 1/|\Omega|^\gamma$ (the $\gamma$-model) and analyze the  competition/interplay between the tendency towards pairing and 
 towards non-Fermi liquid behavior. We show that superconductivity  still develops  once the pairing interaction exceeds a certain threshold, but the origin of the pairing  is qualitatively different from that in BCS theory. We show that the gap equation at $T=0$ has an infinite set of  
   topologically distinct solutions. These solution disappear one by one once the pairing interaction becomes non-singular (massive).   We review the physics underlying these phenomena and outline future directions.
\end{abstract}

\maketitle
\tableofcontents

\section{INTRODUCTION}

Superconductivity remains one of the most spectacular macroscopic quantum phenomena in condensed matter physics.  In conventional metals, the pairing of electrons is elegantly described by the Bardeen-Cooper-Schrieffer  (BCS) theory \cite{Bardeen1957}, later extended by Eliashberg~\cite{Eliashberg1960}. We will use the abbreviation BCS/E  for  BCS/Eliashberg theory.  
 The key narrative of BCS/E 
 is that a phonon-mediated attraction binds fermions into pairs, which almost simultaneously 
  form a macroscopic coherent condensate breaking $U(1)$ gauge symmetry. 
   A cornerstone of this paradigm is an assumption of a  Fermi liquid (FL) behavior in the normal state prior to the onset of superconductivity. This ensures the existence of well-defined quasiparticles close to the Fermi surface, which give rise to a logarithmical enhancement of the pairing susceptibility at low temperatures (the Cooper logarithm). Because of this logarithm, superconductivity develops even if an attraction between Fermi liquid quasiparticles is infinitesimally small.   
   
    The discoveries of strong coupling behavior and unconventional superconductivity in  heavy fermion materials, high-$T_c$ cuprates, iron-based pnictides and, more recently, graphene-based systems and transition metal dichalcogenides~
    \cite{Mathur1998, Taillefer2010,Scalapino2012,Shibauchi2014,Greene2020,cao2018unconventional,cao2018correlated,EAndreiREV2021,
    Pasupathy2024superconductivity,Mak_supercond_2024,zhou2022isospin,zhang2022spin,holleis2023ising}
  called for an understanding 
  of superconductivity in a situation when the strong 
   interaction between fermions leads to largely incoherent behavior
 and the destruction of FL quasiparticles in the normal state above superconducting $T_c$. 
Examples include metals near a spin or charge quantum-critical point (QCP) (see e.g., 
~\cite{Hertz1976, Millis1993,Bergeron2012,Wang_2013,Wang_2025,raghu2015,Torroba_2017,Rohe_2005,Metzner2006,Yamase_2016,ital,ital2,ital3,
acf,acs,acn,Max_2010,Max_2010a,Max_2011,
Max_2015,Lee_2018,Varma2020,Abanov2020,
Shi_2025,Foster_2023, Chowdhury_2013,wang_22}),  Yukawa-SYK models~\cite{Chowdhury2022,Chowdhury2020a,esterlis2026,esterlis2019,HAUCK2020168120,classen2021,wang2020,wang2020solvable} and their extensions ~\cite{Kim2021,Stangier2026,Stangier2026_1}. In these systems,  fermions are coherent at the bare level, but loose coherence in a self-consistent one-loop treatment (this is termed as  "self-tuning to criticality")  and also systems which are 
 not necessary close to a QCP,  yet display incoherent behavior due to numerically strong renormalizations
 ~\cite{Georges1996,simard2019,Capone2009,Capone2023,Held2024,Arovas2022},
The mechanism of superconductivity without coherence is different from that of BCS/E   because fermionic incoherence eliminates the Cooper logarithm. This analysis of this novel mechanism is the prime focus of this review.

Systems displaying a  non-FL behavior and unconventional superconductivity out of  can be broadly separated into two types. 
  In systems of the first type,   an  electronic bandwidth $W$ and a Fermi energy $E_F$  (typically of the order of $W$)  are both larger than the dynamical interaction taken  at characteristic energies, relevant to non-FL and superconductivity. In systems of the second type,   relevant interactions  are larger than the bandwidth.
  Systems of first type include  metals near a $T=0$ instability towards some electronic order: fermions in these systems are incoherent at low energies but still itinerant (see e.g., ~\cite{Hertz1976, Millis1993}.  Systems
  of the second type include Hubbard-type models with Hubbard $U$ larger than $W$.  In these systems,   a localization of electrons (Mott physics) plays the crucial role (see e.g., \cite{Georges1996,simard2019}).  For this  review, we focus on systems of the first type, in which $W$ and $E_F$ are the largest parameters, e.g., itinerant metals near a quantum-critical point (QCP) towards spin or charge order.  We will show that Yukawa-SYK models, formally belonging to the second class,  display nearly identical behavior in a certain limit. 
     The assumption that $W$ and $E_F$ are the largest energy scales also holds in 
     BCS/E theories. Yet, we argue below that the fact that the effective interaction is (i) dynamical and (ii) singular  changes the  pairing mechanism entirely and gives rise to a number of  qualitative differences with  BCS/E.  
     
In this review we focus on six  such differences:
\begin{itemize}
\item 
{\bf Hierarchy of scales}. ~~ 
 In BCS/E theory, the largest energy scale is the effective mass of a boson, $\Lambda$ (a Debye frequency for an Einstein phonon), while  $T_{c}$ and superconducting gap  $\Delta $ are much smaller.  For systems with singular  dynamical interaction,  $\Lambda $ is the smallest scale in the problem, while $T_{c}$ and $\Delta $ are much larger. 
\item  {\bf Universality.}~~ In BCS/E theory, $T_c$ and $\Delta$ depend  on three parameters: the dimensionless coupling in the particle-particle channel, the dimensionless coupling in the particle-hole channel, which gives rise to mass renormalization,  and $\Lambda$. 
    (the ratio $\Delta/T_c$  
 is independent of 
$\Lambda$).  The two dimensionless couplings are
 generally different but 
 comparable in strength and each  
  can be viewed as the ratio 
      of the effective interaction ${\bar g}$ (defined below and assumed to be smaller than $W$ ($E_F$)) and $\Lambda$.  When $\Lambda$ is the smallest energy scale, 
 ${\bar g}$  becomes the single relevant  energy scale, and when all quantities are measured in units of ${\bar g}$, the analysis of a non-FL and  pairing for a given singular interaction becomes a universal problem, in which the only dimensionless parameter (a number)  is the ratio of the interactions in the particle-particle and particle-hole channels.  We will label this  number as $1/N$ below. 
    \item {\bf Threshold.}~~ In BCS/E theory (large $\Lambda$)  superconductivity develops  already for arbitrary attractive interaction in the particle-particle channel, even if it is much weaker than the interaction in the particle-hole channel, i.e., in our notations, for arbitrary large $N$. This is the consequence of the Cooper logarithm. In the opposite limit of vanishingly small $\Lambda$, there exists a threshold on $N$ ($N = N_{cr}$), below which (at $N > N_{cr}$), the ground state remains a non-FL 
        \item {\bf Gap structure.}~~ In BCS/E theory,  the pairing gap $\Delta (\omega)$ is essentially a  constant up to the frequencies of order $\Lambda$.  At small $\Lambda$, $\Delta (\omega)$ is a universal function of frequency when both are measured in units of ${\bar g}$. In other words, the gap evolves at frequencies comparable to its magnitude at $\omega =0$.
    \item {\bf Number of solutions for the gap.}~~ 
    In BCS/E theory, there  is a single solution of the gap equation.  At vanishingly small $\Lambda$, 
    there is an infinite number of solutions. As $\Lambda$ increases, the solutions disappear one by one. 
\item {\bf Topology.}~~The gap function in BCE/E theory has no vortices in the upper (causal) half-plane of complex frequency. In 
contrast, 
at $\Lambda \to 0$, the gap functions from the 
infinite set
       are all topologically distinct. The corresponding topological number  
       is the number of zeros of $\Delta (z )$ in the upper half-plane of $z = \omega' + i \omega''$, or the winding number of the complex function $\Delta (\omega )$ on the real frequency axis.
\end{itemize}

\section{SINGULAR DYNAMICAL INTERACTION }\label{sec:strong}

We consider superconductivity in a situation when above $T_c$ a 
 coherent motion of low-energy fermionic quasiparticles  with momenta near the Fermi surface 
 is destroyed. This destruction comes from a dynamical self-energy $\Sigma (k,\omega)$ 
  whose frequency dependence at $\omega > \Lambda$ must be stronger than in a Fermi liquid. 
 The strong  dynamical self-energy originates from strong  
singular dynamical interaction in the particle-hole channel~\cite{agd}: 
$ \Gamma_{ph} (({\bf k}_F, \omega_k), ({\bf p}_F, \omega_p); ({\bf p}_F, \omega_p), ({\bf k}_F, \omega_k)) = \Gamma_{ph} ({\bf k}_F - {\bf p}_F,  \omega_k - \omega_p)$. 
          We will see below that for the set of problems that we consider, the self-energy is expressed via 
        $\Gamma_{ph}$ integrated along the Fermi surface:    
 \beq
   {\bar \Gamma}_{ph} (\omega_k - \omega_p) = \oint \oint d {\bf k}_F  d {\bf p}_F  \Gamma_{ph} ({\bf k}_F - {\bf p}_F,  \omega_k - \omega_p) 
   \label{a_1}
   \eeq 
The same dynamical interaction viewed in the  particle-particle channel is an effective pairing interaction between fermions: $\Gamma_{pp} ((k, \omega_k), (-k, -\omega_k); (p, \omega_p), (-p, -\omega_p)) = \Gamma_{pp} ({\bf k}_F - {\bf p}_F,  \omega_k - \omega_p)$. 
         It  is not always attractive, but in almost all systems studied so far, there exists at least one 
   attractive pairing channel.  We focus on this  channel and  define the corresponding eigenfunction as $\eta_k$.\footnote{For $d_{x^2-y^2}$ pairing, $\eta_k = (\cos k_x - \cos k_y) f({\bf k})$, where $f({\bf k})$ has full lattice $D_{4h}$ symmetry. For hole-doped cuprates, $f({\bf k})$ was argued to be well approximated by a constant in a wide range of dopings~\cite{Ramshaw2026}. }  
 The pairing interaction in this channel is
        $\Gamma_{pp}$, weighed with $\eta_k \eta_p$ and integrated along the Fermi surface:
    \beq
   {\bar \Gamma}_{pp} (\omega_k - \omega_p) = \oint \oint d {\bf k}_F  d {\bf p}_F \eta_k \eta_p  \Gamma_{pp} ({\bf k}_F - {\bf p}_F,  \omega_k - \omega_p) 
   \label{a_2}
   \eeq  
   The   dynamical interactions  $ {\bar \Gamma}_{ph} (\Omega)$ and $ {\bar \Gamma}_{pp} (\Omega)$ 
   are not identical for non-ordinary $s$-wave pairing as the pairing component has been integrated with the  momentum eigenfunction. However, they show similar behavior and to the first approximation can be treated as proportional to each other, differing by $1/N$, which we introduced earlier.   For these interactions,  $\Lambda$ is the scale below which 
     ${\bar \Gamma}_{ph} (\Omega)$ and  ${\bar \Gamma}_{pp} (\Omega)$ can be  approximated as frequency independent.   
     
     These two interactions have been explored numerous times at large $\Lambda$,  within the  Eiashberg formalism for  electron-phonon systems and for non-critical metals with  4-fermion interaction  mediated by fluctuations in spin or charge channel.  In this case, 
 frequency dependencies  of 
     ${\bar \Gamma}_{ph} (\Omega)$  and  ${\bar \Gamma}_{ph} (\Omega)$  can be largely neglected. The only exception is the computation of $T_c$ in  which  $\Lambda$  acts the upper cutoff for the Cooper logarithm, and using the full
      form of $\Gamma_{pp} (\Omega)$ allows one to 
     compute $T_c$ with the exact prefactor~\cite{Eliashberg1960}.  
        As we said, the situation changes qualitatively when $\Lambda$ becomes small, the hierarchy of the energy scales changes,  fermionic coherence in the normal state is lost and the tendency towards pairing competes with the tendency towards non-FL.  Below we  summarize the results for  the most extreme case when 
        $\Lambda =0$.   Because ${\bar \Gamma}_{ph} (\Omega)$  and  ${\bar \Gamma}_{ph} (\Omega)$ naturally decrease with increasing $\Omega$, the vanishing of $\Lambda$ implies that ${\bar \Gamma}_{ph} (\Omega)$  and  ${\bar \Gamma}_{ph} (\Omega)$  become singular functions of frequency.  We model these singular interactions by power-law dependence specified by the exponent $\gamma$ (the $\gamma-$model, see. e.g.,  \cite{Abanov2020}): 
  \beq
 {\bar \Gamma}_{ph} (\Omega_m) = \left(\frac{\bar g}{|\Omega_m|}\right)^\gamma, ~~{\bar \Gamma}_{pp} (\Omega_m) = \frac{1}{N} \left(\frac{\bar g}{|\Omega_m|}\right)^\gamma,    
   \label{a_3}
   \eeq    
We list the values of $\gamma$ for specific models  in Sec. \ref{sec:models}.  
 The  theoretical analysis  differs somewhat between $\gamma <1$ and $\gamma >1$. In what follows we focus on 
 $\gamma <1$.  The analysis for $\gamma \geq 1$ has been presented in  References \cite{paper3,Wu2021,paper5,paper6,esterlis2026,esterlis2019,HAUCK2020168120}.

We show that for $\gamma <1$,  there exists a critical $N_{cr} >1$, above which superconductivity does not develop and the system  remains in the non-FL normal state down to $T=0$.  At $N < N_{cr}$, superconductivity does develop, but for  a reason fundamentally different from BCS/E: Cooper logarithm is irrelevant and the instability towards pairing is indicated by the appearance of a complex exponent for the pairing susceptibility. 
   We show that this non-BCS/E  pairing gives rise to an infinite number of solutions $\Delta_n (\omega_m)$  of the non-linear gap equation (there is only one solution in BCS/E theory). 
   The solutions  are specified by integer $n$ and are topologically distinct in the sense that  $\Delta_n (\omega_m)$ 
   has $n$ zeros on the upper half of the Matsubara axis. We show that each zero is a dynamical vortex, which gives rise to $2\pi$ phase slip on the real axis.  For the pure $\gamma$-model 
    the global energy minimum is for topologically trivial solution $n=0$, while the  solutions with a finite $n$ are  saddle points with $n$ unstable directions towards the solutions with smaller $n$.   However, for a more complex model,  a vortex solution with $n >0$ becomes the ground state in a certain parameter range
    ~\cite{yu2026}.
       We also argue that even the $n=0$ solution is not BCS/E-like:
     $T_c$  has a power-law rather than exponential dependence of the coupling constant 
      the  ratio of $2\Delta (0)/T_c$  depends on $\gamma$, and the feedback on fermions from the pairing is fundamentally different from that in BCS theory.

\begin{figure}[h]
\centering
\includegraphics[width=\textwidth]{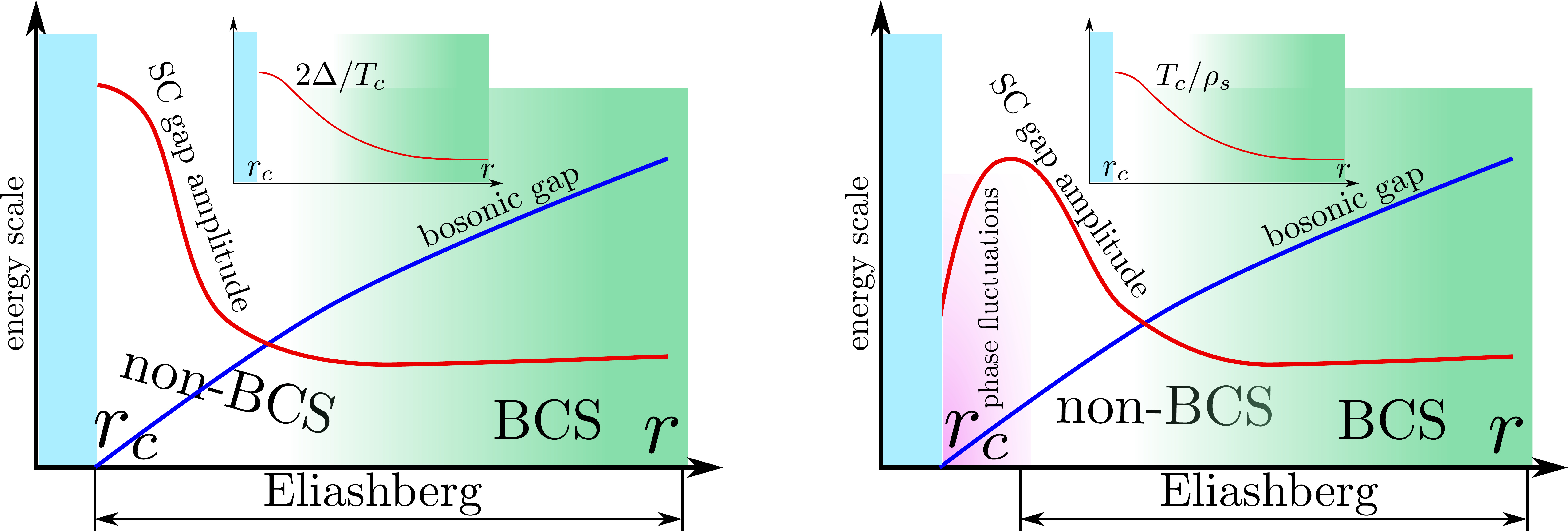}
\caption{Schematic representation of the evolution of superconductivity behavior between FL and non-FL regimes. 
  For metals with itinerant electrons,  parameter  $r$ measures the distance to a QCP at $r_c$. For other systems 
   we treat $r-r_c$ as a proxy to $\Lambda$, which is energy above which the dynamical component of interaction cannot be neglected. Panel (a) refers to systems in which vertex corrections remain small down to $r=r_c$, where $\Lambda =0$. These systems can be analyzed within self-consistent one-loop approximation (a generalized Eliashberg theory).
    Panel (b) is for systems in which vertex corrections become large at small enough $r-r_c$. In these systems,  Eliashberg description holds only at some deviation from $r_c$. Superfluid stiffness $\rho_s$ becomes comparable to $T_c$ at the lower boundary  for Eliashberg theory, and at smaller $r-r_c$, $T_c$ is constrained by phase fluctuations}
\label{fig:BCSnonBCS}
\end{figure}

Our results are summarized in \textbf{Figure \ref{fig:BCSnonBCS}}, where we used a metal near a QCP as an example. 
The parameter $r$ sets the distance to a QCP at $r = r_c$ and $\Lambda$ scales with  $r-r_c$ and vanishes at $r=r_c$.
 For other examples of strong dynamic interaction, not necessary related  to a QCP, one can just assume that $\Lambda$ is small enough.  We argue that as  $\Lambda$ becomes smaller than $T_c$ (modulo a numerical factor)  the system crosses over from a conventional BCS regime to a non-BCS regime, where the pairing still holds, if $N < N_{cr}$ in  Equation \ref{a_3}, but the pairing mechanism is qualitatively different from BCS.  We argue below that over some range of $\Lambda$ 
  this non-BCS pairing can be described within the framework of a modified Eliashberg theory, which includes non-FL 
  self-energy but neglects vertex corrections.  Depending on the model, this range either extends down to $\Lambda =0$ (left panel in \textbf{Figure \ref{fig:BCSnonBCS}}) or down to some small but finite $\Lambda$, and at even smaller
   $\Lambda$ vertex corrections become essential (right panel).  In this last regime, phase fluctuations become strong and $T_c$ is determined by superfluid stiffness $\rho_s$ and likely goes down at $r \to r_c$.
    For the $\gamma$ model,  prototypical examples for left and right  panels are the cases $\gamma <1$ and $\gamma >1$, respectively.

   \section{BCS PARADIGM, COOPER LOGARITHM AND ELIASHBERG THEORY}

BCS theory~\cite{Bardeen1957,Eliashberg1960,agd}  has been discussed multiple times in the literature and we will be brief and avoid the details.   Because $W$ and $E_F$ are assumed to be much larger than $T_c$ and $\Delta$,    
   the fermionic dispersion $\xi_{\mathbf{k}}$ can be linearized in the vicinity of the Fermi surface as $\xi_{\mathbf{k}} \approx v_F(k - k_F)$, which simplifies the analysis of the pairing kernel.

In BCS/E theory, ${\bar \Gamma}_{ph} \equiv \lambda$ and ${\bar \Gamma}_{pp} \equiv \lambda/N$ 
are treated as frequency independent (instantaneous) up to $\Lambda$, which here acts as a sharp cutoff. 
 By order of magnitude, $\lambda \sim {\bar g}/\Lambda$, where ${\bar g}$ is roughly the same as in Equation \ref{a_3}.  
 The BCS/E pairing instability  can be obtained in various ways. For comparison with non-BCS case later in the next section, we introduce a bare infinitesimal gap function $\Delta_0$ and compute the pairing susceptibility at  temperature $T$,  $\chi_{pp} (T)$,  as the ratio of fully renormalized and bare gap functions: $\chi_{pp} (T) = \Delta (T)/\Delta_0$.   Mathematically, this amounts to  summing up series of ladder diagrams in the particle-particle channel. This is most easily done in Matsubara representation, where the fermionic frequencies are discrete $\omega_m = \pi T (2m+1)$. 
    Each cross section contains the Cooper logarithm: 
    $\pi T \sum_{|\omega_m| < \Lambda} 1/|\omega_m| = \log{\frac{1.13 \Lambda}{T}}$. 
   Because the fermions are coherent, the summation generates a perfect geometric series:
\begin{equation}
\chi_{pp} (T) = 1 + \frac{1}{N} \frac{\lambda}{1+ \lambda} \log\frac{1.13 \Lambda}{T} + \left(
\frac{1}{N} \frac{\lambda}{1+ \lambda} \right)^2 \log^2\frac{1.13 \Lambda}{T} + \dots
\label{eq:chi_bcs}
\end{equation}
This geometric series trivially sum up into $\chi_{pp} =  \left(1 - \left(\frac{1}{N} \frac{\lambda}{1+ \lambda}\right) \log\frac{1.13\Lambda}{T}\right)^{-1}$.  The pairing susceptibility diverges at $T= T_c = 1.13 \Lambda 
 e^{-N (1+ \lambda) /\lambda}$ indicating that a superconducting instability develops for arbitrary large $N$, i.e., even for infinitesimally small ${\bar \Gamma}_{pp}$.     Performing similar calculation at $T=0$  one obtains a non-zero $\Delta (T=0) = \Delta = a T_c$.  The prefactor $a$ depends somewhat  on the symmetry of the pairing state (e.g., 
$a = 1.78$ 
for s-wave pairing,  $a = 2.15$ for $d-$wave pairing), but for order of magnitude estimates we can just set $\Delta \sim T_c$.  
 
  Three outcomes of this consideration are essential for the comparison with the analysis of non-BCS pairing below:   
  (i) $T_c$ and $\Delta$ are non-zero for all $N$, no matter how large,  (ii) both are exponentially small in $1/\lambda$  when $\lambda$ is small;  this sets the hierarchy $T_c, \Delta \ll \Lambda$, and (iii)  
 the number of  Matsubara  frequencies involved in the calculation of $T_c$ and $\Delta$  is exponentially large, of order $\frac{\Lambda}{2\pi T_c} \sim e^{N (1+ \lambda)/\lambda}$. In this situation,  the result of frequency summation is  insensitive to the behavior at any given frequency.
 
 \subsection{Eliashberg Theory}

The canonical Eliashberg theory~\cite{Eliashberg1960} is an extension of BCS theory to energies above $\Lambda$, where the frequency dependence of the interaction becomes relevant.     The theory 
 has been originally  developed for phonon-mediated  $s-$wave superconductivity ~\cite{Eliashberg1960}, but has been later extended to the cases  when the pairing is mediated by gapped collective excitations in spin or charge channel.    
    
    Two aspects of Eliashberg theory are essential to our analysis of non-BCS pairing below.  First, even at small $\lambda$, it allows one to     
 compute $T_c$ and $\Delta$ explicitly rather than  expressing it in terms  $\Lambda$, which is defined in order of magnitude estimates.  The result depends on the precise form of the effective 4-interaction.   
 Second and most relevant,  Eliashberg theory allows one to extend BCS analysis to $\lambda \geq 1$, when the full dynamical self-energy is relevant and the feedback from superconductivity on fermions is essential. Still, the canonical Eliashberg theory assumes that both $T_c$ and $\Delta$ are much smaller  than $W$ ($E_F$).   

    Eliashberg theory for $\lambda \geq 1$ consists of the set of two coupled integral equations on fermionic self-energy and the pairing vertex. They can be re-expressed as a single integral equation on the gap function and the equation for the self-energy in terms of the gap function.   These equations are obtained within a  self-consistent one-loop approximation, by including series of renormalizations of propagators of internal fermions but neglecting vertex corrections. These corrections generally hold in powers of $ \lambda \Lambda/E_F$.  For electron-phonon interaction, $\Lambda$ is the Debye frequency $\omega_D$,  and vertex corrections hold in powers of Migdal-Eliashberg parameter $ \lambda_E = \lambda \omega_D/E_F$ References \cite{Migdal1958, Eliashberg1960}. This parameter is small in 
the adiabatic approximation when lattice vibrations are slow modes in comparison to electrons.  For pairing mediated by massive  collective excitations, $\lambda_E \sim {\bar g}/E_F$, which is also often  small, at least numerically.   
    The smallness of $\lambda_E$ allows one to also  neglect the  Landau damping of a phonon and its modification in the superconducting state and factorize the momentum integration in the diagrammatic series for the self-energy and the pairing vertex by 
    integrating  over the momentum perpendicular to the Fermi surface  only in fermionic propagators,  the  remaining  integration over the momentum along the Fermi surface is in the bosonic propagator, between the momenta on the Fermi surface~(see e.g., \cite{Haslinger2003}).  Solving Eliashberg equations for $\lambda  \geq 1$ gives a highly reliable value of $T_c$ and is  frequently  used to predict superconductivity in  moderately strongly coupled materials.   
    
     In what follows, we extend the Eliashberg approach to the limit  $\Lambda \to 0$ and $\lambda \to \infty$  keeping $\lambda_{E}$  small and show that 
        superconductivity in this limit is qualitatively different from the  one at a finite $\lambda$.  When $\lambda$ is small but finite, there is a crossover between 
          the conventional behavior  and the new one.  Because $T_c$ is finite even when $\lambda$ is infinite,  the crossover to the conventional behavior occurs only when  $1/\lambda$  exceeds a certain  threshold. .

\section{NON-BCS PAIRING BY SINGULAR DYNAMICAL INTERACTION}

\subsection{Models} 
\label{sec:models}

We now consider pairing in the situation when $\Lambda$ is smaller than the energy scale associated with superconductivity.   Two models have been extensively analyzed in the context of pairing in this situation. One is the $\gamma$-model of a quantum-critical metal,  another is the Yukawa-SYK  model. We introduce them below and show that they yield identical gap equations at $T=0$. We then continue with the $\gamma$-model.   Non-FL and pairing in the Yukawa-SYK model has been reviewed in~\cite{annurevSchmalian}. 
     
\subsubsection{Metal near  a Quantum Critical Point}

At a QCP in a metal,  the dominant interaction between fermions is via the exchange of dynamical fluctuations of an 
 order parameter which condenses on the ordered side of a QCP.  In the situation when the momentum integration in diagrammatic series for the self-energy and the pairing vertex can be factorized (the case that we consider), this interaction is proportional to the dynamical susceptibility of the order parameter, averaged over momenta taken between points on the Fermi surface.   Order parameter fluctuations are massless at a QCP, and the momentum-averaged propagator has a singular  dependence on frequency, which channels into a singular effective interaction in Equation~\ref{a_3}, specified by the exponent $\gamma$.  The interaction in Equation~\ref{a_3} decreases at large $\Omega$ for all $\gamma >0$,  which eliminates the need to introduce a high-frequency cutoff. 
  Microscopic models studied in this context  include pairing in 
 $3D$  
 systems, where $\gamma = 0_+$ (the same exponent for color superconductivity~\cite{son,son2}),
    spin- and charge-mediated pairing in $D=3-\epsilon$ dimensions~\cite{senthil,Max_2015,Fitzpatrick_2015} (and in graphene near charge neutrality~\cite{khveshchenko}), where 
 $\gamma = O(\epsilon)$,
2D systems at the onset of  $2k_F$ order,  where  $\gamma =1/4$ (Reference ~\cite{2kf}),   systems near the onset of  nematic/Ising-ferromagnetic order~\cite{nick_b,steve_sam,triplet,triplet2,triplet3,Klein_2018,Klein_2020,Liu_2022}, where $\gamma =1/3$ (the same exponent as for the interaction mediated by a transverse gauge field~\cite{Lee1992}), systems near the onset of 
     CDW and SDW order in 2D ~\cite{Bergeron2012,Wang_2013,Wang_2025,raghu2015,Torroba_2017,Rohe_2005,Metzner2006,Yamase_2016,ital,ital2,ital3,
     acf,acs,acn,
     Max_2010,Max_2010a,Max_2011,Lee_2018,Abanov2020}, for which $\gamma =1/2$, a 2D pairing  mediated by an undamped  propagating boson ($\gamma =1$),  pairing in several Fe-based superconductors~\cite{kotliar} ($\gamma =1.2$), pairing near a relativistic Mott transition in twisted Dirac materials
     \cite{Stangier2026} ($1<\gamma<2$),   and 
  the strong coupling limit of phonon-mediated superconductivity ~\cite{combescot,Bergmann,Bergmann2,ad,Marsiglio_88,Marsiglio_91,Karakozov_91} ($\gamma =2$).  The pairing models with parameter-dependent $\gamma$ have been analyzed as well (Reference \cite{Subir,moon_2}).  
  
 The full set of Eliashberg-type equations for the pairing at a QCP ($\Lambda =0$) consists of a set of  coupled equations for the fermionic self-energy, the pairing vertex, and the bosonic polarization bubble. The latter is  relevant in 
 the
non-linear regime below $T_c$ as fermionic  pairing affects the Landau damping part of the 
 propagator of the order parameter.   For simplicity of presentation, we skip this feedback effect and 
   restrict with the set of  two coupled equations on the Matsubara axis for the self-energy  $\Sigma (\omega_m)$ and the pairing vertex $\Phi (\omega_m)$.  The gap function $\Delta (\omega_m)$ is related to $\Phi$ via
 \beq
\Delta(\omega_m) = \Phi(\omega_m) \frac{\omega_m}{\omega_m + \Sigma(\omega_m)}.
\eeq
    We rescale $\omega_m$, $\Sigma$ and $\Phi$  by ${\bar g}$ from Equation~\ref{a_3}, i.e., make them dimensionless. 
      We will be using a dimensionless $\omega_m \equiv \omega_m/{\bar g}$ below. 
      In  dimensionless  variables, the  equations for  $\Sigma$ and $\Phi$   become completely universal with numbers $\gamma$ and $N$  as parameters: 

  \begin{equation}
\Sigma(\omega_m) = \pi T \sum_{\omega_n} \frac{\Sigma(\omega_n)}{\left[(\omega_n + \Sigma(\omega_n))^2 +\Phi^2 (\omega_n)\right]^{1/2}}
 \frac{1}{|\omega_m - \omega_n|^\gamma}
\label{eq:sigma_eliashberg}
\end{equation}
\begin{equation}
\Phi(\omega_m) = \frac{\pi T }{N} \sum_{\omega_n} \frac{\Phi(\omega_n)}
{\left[(\omega_n + \Sigma(\omega_n))^2 +\Phi^2 (\omega_n)\right]^{1/2}}
 \frac{1}{|\omega_m - \omega_n|^\gamma} 
\label{eq:phi_eliashberg}
\end{equation}
It is customary to call Equation~\ref{eq:phi_eliashberg} the  gap equation.
In the normal state, the solution of Equation~\ref{eq:sigma_eliashberg} for the self-energy is 
\begin{equation}
\Sigma(\omega_m) = \frac{1}{1-\gamma} |\omega_m|^{1-\gamma} \text{sgn}(\omega_m)
\label{eq:sigma_nfl}
\end{equation}
This fractional power-law dependence has a profound implication: at low frequencies, the self-energy is 
parametrically larger than the bare Matsubara frequency.  Converting Equation~\ref{eq:sigma_nfl} to real frequencies, we 
 obtain $\Sigma'(\omega) \sim \Sigma'' (\omega) \sim \omega^{1-\gamma}$.  This implies that the Landau criterion for a Fermi liquid behavior $\Sigma'' (\omega) \ll \omega$ is not satisfied.  By definition, in this case the system displays a non-FL behavior down to $\omega =0$. 
 In physical terms,  the fermions cease to be  propagating quasiparticles and become completely incoherent.
     
Using  Equation~\ref{eq:sigma_nfl}, one can explicitly write the "linearized" equation for an infinitesimally small pairing vertex, which allows one to determine whether the 
normal state is unstable towards pairing.  We have  
\begin{equation}
\Phi(\omega_m) = \pi T \frac{1-\gamma}{N} 
\sum_{\omega_n} \frac{\Phi(\omega_n)}
{|\omega_n|^{1-\gamma} + (1-\gamma) |\omega_n|}
 \frac{1}{|\omega_m - \omega_n|^\gamma} 
\label{eq:phi_eliashberg_l}
\end{equation}
 At $T=0$, $\pi T \sum_{\omega_n}$ is replaced by $\tfrac{1}{2} \int d \omega_m$ and the linearized gap equation becomes 
 \begin{equation}
\Phi(\omega_m) =\frac{1-\gamma}{2N} 
\int\! d \omega_n  \frac{\Phi(\omega_n)}
{|\omega_n|^{1-\gamma} + (1-\gamma) |\omega_n|}
 \frac{1}{|\omega_m - \omega_n|^\gamma} 
\label{eq:phi_eliashberg_l_1}
\end{equation}
  We will use this equation for the calculation of the pairing susceptibility in Sec. \ref{sec:pairing_chi}.

 \subsection{Yukawa SYK model}

Yukawa SYK model is  a generalization of the Sachdev-Ye-Kitaev (SYK) model \cite{PhysRevLett.70.3339,Kitaevtalk,PhysRevLett.85.840,PhysRevLett.105.151602,Kitaev2018}
It describes ${\tilde N}$  dispersion-less fermions in a quantum dot 
randomly coupled by a complex Yikawa-type interaction to $M$ Einstein phonons with a finite bare Debye frequency $\omega_D$~\cite{Chowdhury2022,Chowdhury2020a,esterlis2026,esterlis2019,HAUCK2020168120,classen2021,wang2020,wang2020solvable}. 
The coupling between electrons and bosons is responsible for incoherent NFL behavior \emph{and} electronic pairing. 
 An appeal of the Yukawa-SYK model is that 
 it becomes
exactly solvable in the limit ${\tilde N}, M \to \infty$, ${\tilde N}/M$ finite.     In this limit, vertex corrections vanish and self-consistent one loop approximation becomes exact. 
 The highly non-trivial feature  of the model is a self-tuning to
  quantum criticality  for any value of 
 $\omega_D$:  in self-consistent treatment the fully renormalized $\omega_D$ vanishes.
   Like in a quantum-critical metal, there is a competition  between NFL behavior and pairing.
   The outcome depends on  ${\tilde N}/M$, and on the ratio of real and imaginary components  of the Yukawa coupling, which 
      plays the same role as $1/N$ in quantum-critical models. 
    The analogy with a QC metal goes further as once vertex corrections are neglected,  Yukawa-SYK model can be  equivalently viewed as the one with two non-random singular dynamical four-fermion interactions in particle-hole and particle-particle channels with relative strength $1/N$ and the same power-law structure as in Equation~\ref{a_3}
     with exponent $\gamma$ expressed via ${\tilde N}/M$  (Equation~\ref{eq:eta} below). 
      The interplay between NFL and superconductivity is then again described by a set of coupled equations for the fermionic self-energy, the pairing  vertex, and the bosonic polarization bubble. The latter accounts for self-tuning to criticality.
    
    Eliashberg equations for the Yukaws-SYK model have been derived in References 
    \cite{esterlis2026,esterlis2019,HAUCK2020168120,classen2021,wang2020,wang2020solvable,Stangier2026,Stangier2026_1}.
 They are similar, but not identical to those in a QC metal because of the  lack of fermionic dispersion  in  the  Yukawa-SYK model.  Yet, the linearized equation on $\Phi$ is essentially the same as in a QC metal. 
      Specifically, in the would be normal state at $T=0$ , the fermionic self-energy has a non-FL form
     \begin{align}
 \Sigma (\omega) =\frac{1}{1-\gamma} {\text {sign}}(\omega)|{\omega}|^{(1-\gamma)/2}
 \label{eq:Sigmaqu}
 \end{align}
 where $\gamma$ is  related to ${\tilde N}/M$ as
 \be
\frac{\tilde N}{M}=
\frac{\gamma}{(1-\gamma)^{1/2}}\frac{\tan\left(\pi\frac{\gamma}{2}\right)}{\tan\left(\pi\frac{1+\gamma}{4}\right)}\,.
\label{eq:eta}
\ee
and the self-energy and frequency are again expressed in units of ${\bar g}$, which is made out of system parameters and is the single energy scale. For $M\gg {\tilde N}$, $\eta\rightarrow 0$, for  $M\ll {\tilde N}$,  $\eta\rightarrow1$ and for ${\tilde N}=M$, $\eta=0.6815$. 
 The  linearized equation for the pairing vertex is, in the same dimensionless variables is 
     \begin{align}
\label{eq:lingap1}
\Phi(\omega_m)= \frac{1-\gamma}{2 N} \int\!d\omega_n\frac{\Phi(\omega_n)}{|{\omega_m-\omega_n}|^\gamma
\left(|{\omega_n}|^{(1-\gamma)/2} + (1-\gamma)^{1/2}|\omega_n|\right)^2}
\end{align}
 Comparing with Equation~\ref{eq:phi_eliashberg_l_1}, we see that the two equations for the pairing vertex are essentially  identical. They become completely identical if  we approximate   second, $(a+b)^2$ term in the denominator
  as $a^2 + b^2$. 
  
\subsection{Pairing susceptibility}
\label{sec:pairing_chi}

For a BCS/E superconductor, we identified the pairing instability by analyzing the temperature dependence of the pairing susceptibility.  We now perform the same calculation by adding $\Phi_0$ to the r.h.s. of Equation~\ref{eq:phi_eliashberg_l} and solving iteratively for the full $\Phi$. Due to the dynamical nature of the pairing interaction, the full $\Phi$  now depends on the fermionic frequency $\omega_m$.  For simplicity, we do the calculation at $T=0$.  For the BCS case, $\chi_{pp} (T \to 0)$ is negative as the consequence of its divergence at $T_c$.  We  verify whether in the non-FL case $\chi_{pp} (\omega_m)$ at $T=0$ is negative, at least at some $\omega_m$. If it is, there must be a pairing instability at a finite $T$. If it does not, 
 the normal state remains stable against pairing down to $T=0$.
 
Analyzing the ladder series we note that, at a first glance,  its structure does not fundamentally 
 change compared to the BCS/E case. Indeed, the  series still contains powers of logarithms.   
 This arises because at small frequencies and at $|\omega_m| \ll  |\omega_n|$,  the singular effective interaction reduces to  $1/|\omega_n|^\gamma$ and combines with the singular non-FL self-energy $|\omega_n|^{1-\gamma}$  to produce the  marginal $1/|\omega_n|$ kernel of the gap equation~\ref{eq:phi_eliashberg_l_1} (the same  holds for  Equation~\eqref{eq:lingap1}).  However, the  ladder series are not geometrical as only the upper cutoff of the logarithm is universally set at $\omega_n  \sim  \omega_0 = (1-\gamma)^{-1/\gamma}$, where $|\omega_n|^{1-\gamma} = (1-\gamma) |\omega_n|$,  the lower cutoff  in a given cross-section
  is the running frequency in the subsequent cross-section~\cite{Abanov2024}. 
  Isolating the highest power of the logarithm at each order of iteration we find a series of the form:
\begin{equation}
\chi_{pp}(\omega_{m}) = 1 + \frac{1-\gamma}{N}\log\frac{\omega_0}{|\omega_m|} + \frac{1}{2}\left(\frac{1-\gamma}{N}\log\frac{\omega_0}{|\omega_m|}\right)^2 + \frac{1}{6}\left(\frac{1-\gamma}{N}\log\frac{\omega_0}{|\omega_m|}\right)^3 + \dots
\label{eq:chi_gamma_series}
\end{equation}
The coefficients at each order are the binomial factors of the Taylor expansion of the exponent. Summing up this series explicitly, we obtain 
\begin{equation}
\chi_{pp}(\omega_{m}) = e^{\frac{1-\gamma}{N}\log\frac{\omega_0}{|\omega_m|}} \sim  \left(\frac{|\omega_m|}{\omega_{0}} \right)^{-\frac{1-\gamma}{N}}
\label{eq:chi_gamma_resum}
\end{equation}
We see that the  pairing susceptibility increases at low frequencies but remains  positive and finite for any non-zero  $\omega_m$, in shark contrast with  Equation~\ref{eq:chi_bcs} for the BCS/E case. 

This fundamental discrepancy in logarithmic series can be elegantly understood through a renormalization group (RG) analysis of the 4-fermion pairing interaction \cite{raghu2015,Torroba_2017,Abanov2024}. By analyzing the ladder renormalizations, one can derive the one-loop RG equations for the running coupling $g(L)$. In the standard BCS theory, where $L = \log(\Lambda/T)$ and $g_0 = \lambda/N (1+ \lambda)$, the cross-section with the largest intermediate $L$ 
 is assumed to be somewhere  inside  the ladder series, i.e., there are there are cross-sections on both sides of it .  The renormalizations from the cross-sections on each side of the one with the largest $L$ yield the fully dressed coupling,  and   the non-linear RG equation becomes $dg(L)/dL = g^2(L)$. Solving this strictly recovers the standard Cooper logarithm, $\chi_{pp}(L) = 1/(1 - g_0 L)$. For the QC pairing, where the running scale is $L = \log(\omega_0/|\omega_m|)$ and $g_0 = (1-\gamma)/N$, the cross-section with the largest $L'$ is located at the boundary of the ladder rather than in the middle. As a result, the renormalizations evaluate to $g_0 g(L)$ instead of $g^2(L)$. The one-loop RG equation thus becomes  linear: $dg(L)/dL = g_0 g(L)$. Integrating this linear equation yields the exponential form $\chi_{pp}(L) = e^{g_0 L}  \sim  (1/|\omega_m|)^{(1-\gamma)/N}$. This boundary-dominated RG flow is the mathematical mechanism that transforms the geometric progression into the one that yields an anomalous power law,  averting the logarithmic  singularity at any finite $\omega_m$ \cite{Abanov2024}.

This result leads to a simple conclusion: in a non-FL there is no logarithmic superiority of the particle-particle  channel over the particle-hole channel.  The two have to be treated on equal footings. 
 This is entirely new phenomenon not present in conventional  BCS/E
theory, where the Cooper logarithm 
  ensures that the pairing develops no matter how small the pairing attraction is.  Here, 
   the system either remains in a normal non-FL state down to $T=0$, or  somehow superconductivity develops without the  Cooper logarithm.

  We now recall that in Equation~\ref{eq:chi_gamma_series} we restricted with the highest power of the logarithm at any level of iteration.  This restriction is justified if the numerical prefactor in front of the logarithm is small. 
    In BCS series, the prefactor scales as  $\lambda$ and is certainly small at weak coupling, justifying the logarithmic approximation.  In  
    Equation~\ref{eq:chi_gamma_series}, the prefactor for a logarithm is $(1-\gamma)/N$.  
    Let's  fix $\gamma$ at some value between $0$ and $1$ and vary $N$.   At large $N$, the prefactor is small and the logarithmic approximation is justified. 
      However, at $N = O(1)$, there is no justification to restrict with the leading logarithms in the iterations.
       In this situation we must abandon perturbation theory  and instead try to find  the exact solution for
       $\chi_{pp} (\omega)$ at  $T=0$.

\subsection{Beyond logarithmic approximation: Complex Exponents}

 To proceed, we note that in the infrared limit,  when the dimensionless  $|\omega_m| \ll 1$,
   the singular non-FL self-energy  is parametrically larger than the bare $\omega_m$ and one can safely neglect the latter. The  linearized gap equation at $T=0$ with $\Phi_0$ in the r.h.s. then takes the continuous integral form:
\begin{equation}
\Phi(\omega_m) = \frac{1-\gamma}{2N} \int_{-\infty}^{\infty}  \frac{d\omega_n}{|\omega_m - \omega_n|^\gamma} \frac{\Phi(\omega_n)}{|\omega_n|^{1-\gamma}} + \Phi_0
\label{eq:phi_integral}
\end{equation}
The scale-invariance of the kernel of this equation dictates 
 that $\Phi (\omega_m)$   must be  a pure power-law. We therefore look for a solution that at small $\omega_{m}$ has the form:
\begin{equation}
\Phi(\omega_m) \propto   |\omega_m|^{b -\gamma/2}
\label{eq:phi_ansatz}
\end{equation}
where $b <\gamma/2$ is a dimensionless parameter to be determined. 

Substituting this power-law ansatz  into 
 Equation~\ref{eq:phi_integral}, 
keeping the leading divergent terms at small $\omega_m$ and re-scaling the integration variable by $x = \omega_n/\omega_m$, we find an algebraic equation for the parameter $b$:
\begin{equation}
1 = \frac{1-\gamma}{2N} \int_{-\infty}^{\infty}  \frac{dx}{|1 - x|^\gamma |x|^{1 - \gamma/2 + b}}
\label{eq:b_integral}
\end{equation}
The integral is expressed  in terms of Gamma functions  yielding the relation 
\begin{equation}
N =  \frac{1-\gamma}{2} \frac{\Gamma(\gamma/2 -  b) \Gamma(\gamma/2 +  b)}{\Gamma(\gamma)} \left( 1 + \frac{\cos(\pi  b)}{\cos(\pi \gamma / 2)} \right)
\label{eq:eps_b}
\end{equation}
Analyzing this equation we find that for large $N$, $b = \gamma/2 - (1-\gamma)/N$ such that $\chi_{pp} (\omega_m) \sim 
 (1/|\omega_m|)^{(1-\gamma)/N}$ -- the same as we obtained by summing up the leading logarithms\footnote{Equation~\ref{eq:phi_ansatz}  has another solution with $-b$ instead of $b$, but that one does not satisfy the "boundary" condition at $\omega_m = \omega_0$, see \cite{Abanov2024} for more detail}. 
   Solving Equation~\ref{eq:eps_b} for arbitrary $N$, we find that $b$ deceases with decreasing $N$, and {\it vanishes} at  
\begin{equation}
N = N_{cr} = \frac{1-\gamma}{2} \frac{\Gamma^2(\gamma/2)}{\Gamma(\gamma)} \left( 1 + \frac{1}{\cos(\pi \gamma / 2)} \right)
\label{eq:N_cr}
\end{equation}
For $0\leq \gamma <1$, $N_{cr}>1$.  For $N < N_{cr}$, $b^2$ becomes negative, meaning that $b$ becomes purely imaginary: $b = \pm i\beta$. For imaginary  $b$, the exponent in the power-law ansatz becomes complex and 
 the  pairing susceptibility becomes sign-changing: 
\begin{equation}
\chi_{pp} (\omega_m)  \propto \frac{1}{|\omega|^{\gamma/2}} \cos(\beta  \ln |\omega| + \phi_0)
\label{eq:phi_oscillatory}
\end{equation}
where $\phi_0$ is arbitrary. 
 This log-periodic oscillation implies that for any $N < N_{cr}$, there is an infinite number of ranges of $\omega_m$  where the susceptibility is negative.   This signals that the perturbation theory breaks down.  It is natural to expect that this breaking implies that a non-FL ground state becomes unstable towards superconductivity.  We show below that this is indeed the case. We emphasize that  superconductivity develops not  because of Cooper logarithm
 but through a non-perturbative bifurcation  from real to  complex exponent.  There is both mathematical and physical  analogy between this phenomenon and  ``fall to the center'' phenomenon in quantum mechanics, which appears in the solution of Klein-Gordon and Dirac equations in systems with atomic number $Z> 137/2 = 68.5$ (References \cite{Zeld_Popov,Zohar}). 

\begin{figure}[h]
\centering
\includegraphics[width=\textwidth]{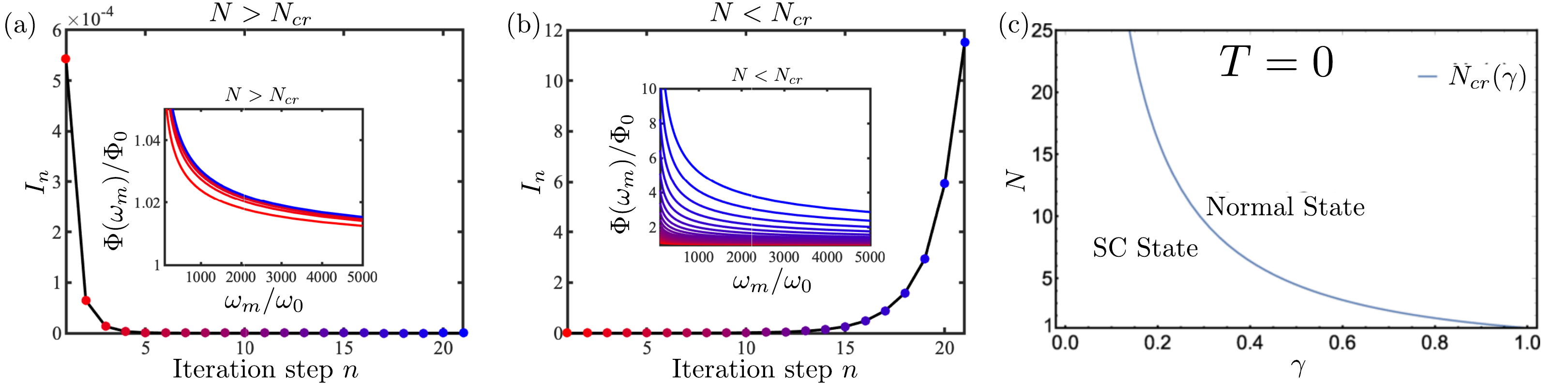}
\caption{The iterative solution of the gap equation with a constant source term $\Phi_0$ for $\gamma = 1/2$. 
The plots show $I_{n}=\frac{1}{2\pi }\int_{0}^{\infty }\left(\Phi^{n+1}(\omega_{m})-\Phi_{n}(\omega_{m}) \right)^{2}d\omega_{m}$. For $N > N_{\text{cr}}$ (a), the iterations converge; for $N < N_{\text{cr}}$ (b), the iterations do not converge. The inserts show how the function $\Phi_{n}(\omega_{m})$ evolves with iterations. For $
N < N_{cr}$ it eventually diverges for all $\omega_m$. (c) The phase diagram with non-FL and superconducting ground states. }
\label{fig:3}
\end{figure}

The full linear  equation for $\Phi (\omega_m)$ at $T=0$ can be actually solved exactly for arbitrary values of $\gamma$ without neglecting the bare Matsubara frequency $\omega$ in comparison with the self-energy $\Sigma(\omega)$ \cite{Abanov2020, Wu2021}. The analytical procedure requires the integral equation to be converted into a Riemann-Hilbert problem.  The large-$\omega$ behavior of $\Phi (\omega_m)$  differs from the scale-invariant case, but  the continuous spectrum of the integral operator remains identical. In particular, 
 the threshold for the onset of pairing, $N_{cr}$, is still given by  Equation~\ref{eq:N_cr}.
    A more easily derivable but still highly accurate solution of the full linear homogeneous equation for $\Phi (\omega_m)$ at $T=0$ can be obtained by converting integral  equation, Equation~\ref{eq:phi_eliashberg_l},  into an approximate differential 
 equation~\cite{Abanov2024} 
by replacing 
  $|\omega_m - \omega_n|^\gamma$ into ${\text {max}} (|\omega_m|, |\omega_n|)^\gamma$.  
  This   approximation can be rigorously  justified  at small $\gamma$ but is numerically quite accurate also at $\gamma \leq 1$ (Reference \cite{Abanov2020}).   
  The solution of the differential equation yields the pairing susceptibility at $N < N_{cr}$ in the form
  \beq
 \chi_{pp} (\omega_m) 
= \frac{H_{i\beta} ( \omega_m) e^{i\phi_0} + H_{-i\beta} (\omega_m) e^{-i\phi_0}}{2 \cos{\phi_0}}
\label{eq:24}
\eeq
 where
 \beq
 H_{i\beta} (\omega_m) = \frac{1+ (1-\gamma) |\omega_m|^\gamma}{|\omega_m|^{\gamma(1/2-i\beta)}}
  \frac{\Gamma\left(\frac{1}{2} +i\beta\right)\Gamma\left(\frac{3}{2} +i\beta\right)}{\Gamma\left(1 +2 i\beta\right) (1-\gamma)^{1/2 -i\beta}} {_2}F_{1} \left[\frac{1}{2} +i \frac{3\beta }{2} +i\beta, 1 +2 i \beta; -(1-\gamma) |\omega_m|^\gamma\right]
 \label{eq:14}
 \eeq
  and ${_2}F{_1} [...]$ is a Hypergeometric function.  We see that for any non-zero $\beta$, there exists a 
  range of  $\phi_0$  (a free parameter in Equation~\ref{eq:24} 
   where $ \chi_{pp} (\omega_m)$ is positive, a range where it is negative, and for a certain
   $\phi_0 = \pi/2$, $\chi_{pp}$ diverges. This non-uniqueness implies that iteration series for $\Phi (\omega_m)$ at $N < N_{cr}$ do not converge for {\it any} $\omega_m$.  In \textbf{Figure \ref{fig:3}} we show the results of our numerical iterative computation of the pairing susceptibility.  We see that $\chi_{pp} (\omega_m)$ remains finite for $N > N_{cr}$ but diverges at all $\omega_m$
  for $N < N_{cr}$, as Equation~\ref{eq:24} indicates. We show in  
 Sec.~\ref{sec:tower}
  that this non-uniqueness of $\chi_{pp}$ is an indicator that  the non-linear gap equation has an infinite number of solutions for  $N < N_{cr}$.
  
 \section{NON-LINEAR GAP EQUATION FOR PAIRING BY SINGULAR DYNAMICAL INTERACTION} 
 
  The solutions of the set of two non-linear Equations~\ref{eq:phi_eliashberg} and \ref{eq:sigma_eliashberg} 
   for a finite $\Phi (\omega_m)$  have been discussed in \cite{Abanov2020}.
  The solutions have been obtained numerically and also analytically, by converting the non-linear integral gap equation into an approximate but highly reliable differential equation. 
  
  The outcome of this analysis is that there exists a single fully stable solution $\Delta_0 (\omega_m)$ -- the absolute minimum of the condensation energy, and a tower of saddle-point solutions $\Delta_n (\omega_m)$ with $n$ unstable  directions in the parameter space. We discuss the  stable solution first and discuss other solution in
   Sec. \ref{sec:tower}.

\begin{figure}[h]
\centering
\includegraphics[width=\textwidth]{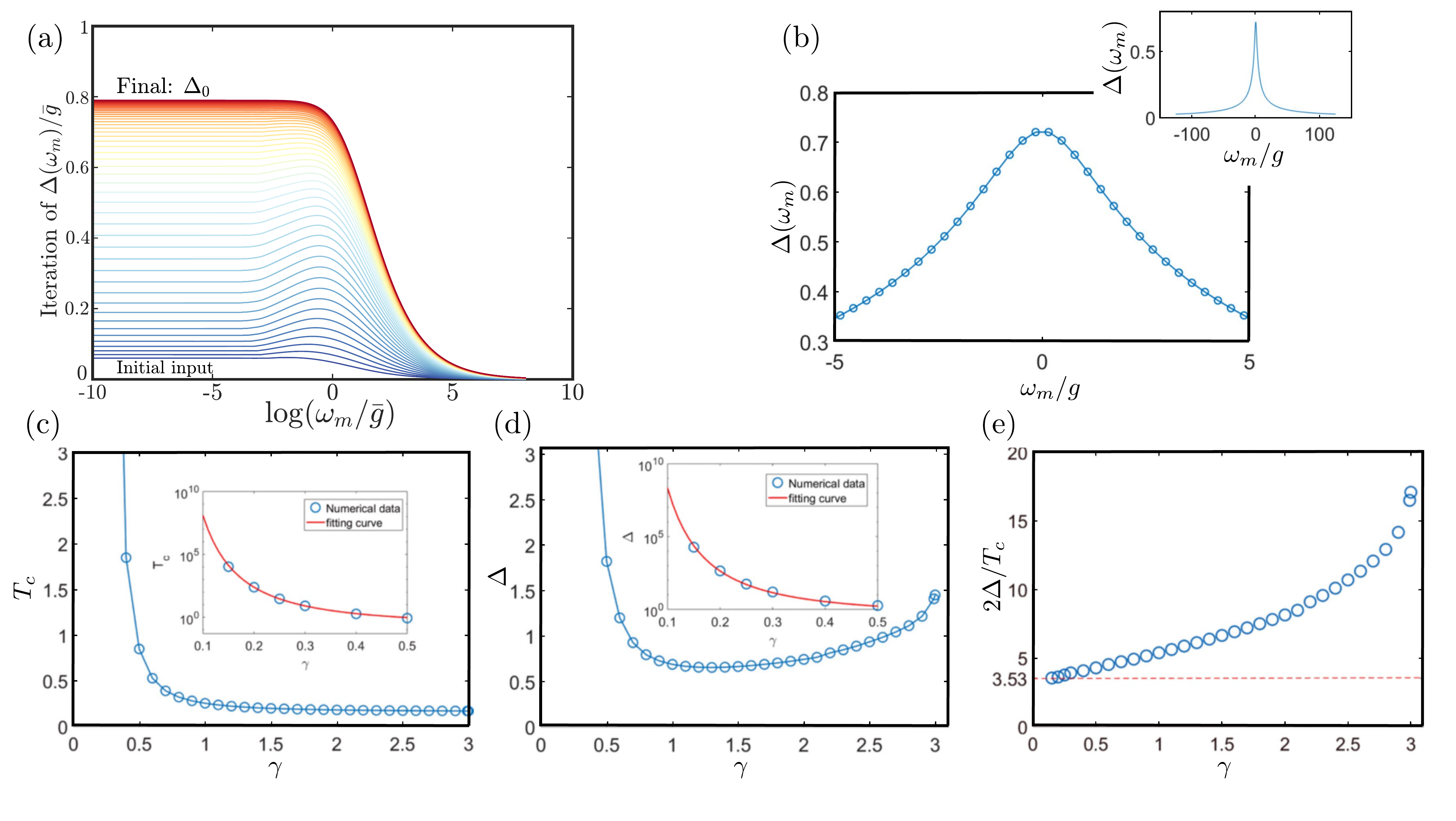}
\caption{Left upper panel: Iteration process for the non-linear gap equation. We depart
  from the infinitesimally small bare $\Delta^{{\text{trial}}}_0 (\omega_m)$  and end up with a finite
    $\Delta_0 (\omega_m)$. Tight upper panel:  the gap $\Delta_0(\omega_m)$ for the stable solution  as a function of the  Matsubara frequency for $\gamma=0.9$, $N=1$ and  
 $T=0.18T_c$. Lower panel: the numerical results for  $T_{c,0}$ (left), $\Delta_0 (0)$  (middle)
  and  the ratio $2\Delta_0 (0)/T_{c,0}$ (right), all as functions of $\gamma$  for $N=1$. }
\label{fig:7}
\end{figure}

  \subsection{Gap structure for $\Delta_0 (\omega_m)$,  $T_c$ and the ratio $2\Delta_0 (0) /T_c$} 
  
  In  the upper panel of \textbf{Figure \ref{fig:7}} we present the results for the iteration process for the non-linear gap equation at $N < N_{cr}$ and the resulting  gap function $\Delta_0 (\omega_m)$ at $T \approx 0$ for the representative case $\gamma =0.9$ and $N =1$. We see that the gap function tends to a finite value at $\omega_m  \to 0$. At large frequencies,  $\Delta (\omega_m)$ decays as $1/ |\omega_m|^\gamma$.  The magnitude of the gap is set by ${\bar g}$, which is the only energy scale for the problem. 
   
 In the lower panel of \textbf{Figure \ref{fig:7}} we show the numerical results for  $T_{c,0}$, $\Delta_0 (0)$ at $T \to 0$  and  the ratio $2\Delta_0 (0)/T_{c,0}$, all as functions of $\gamma$, again for $N=1$. 
  We see that 
  $T_{c,0}$ is just proportional to ${\bar g}$, i.e.,  it does not have an exponential smallness in 
  ${\bar g} \ll E_F$ like in BCS/E theory at weak coupling. 
  The divergence of $T_{c,0}$ and $\Delta_0 (0)$ at $\gamma \to 0$ is artificial as in this limit our problem reduces to the BCS theory without the upper cutoff. If we additionally set the cutoff, $T_{c,0}$ and $\Delta_0 (0)$ will both saturate.  We also see that the ratio $2\Delta_0 (0)/T_{c,0}$ reduces to BCS value  $3.56$ at $\gamma \to 0$, but increases with $\gamma$  towards a larger value and actually diverges at $\gamma \to 3$.  A large $2\Delta_0/T_{c,0}$ 
   for $\gamma =2$ has been noticed before (see e.g., \cite{combescot,Scalapino2012}). 
   
Two observations are relevant here. First, 
thermal fluctuations, described by the terms with $\omega_m = \omega_n$ in Equation~\ref{eq:phi_eliashberg} and Equation~\ref{eq:sigma_eliashberg}, are  formally divergent at a 
  finite $T$, when one has to sum over discrete Matsubara frequencies.
    However, these terms act in the same way as  
    non-magnetic impurities and,  like them,  do not affect $T_c$. These terms cancel out in the equation for $\Delta$ and can be just eliminated from Equation~\ref{eq:phi_eliashberg} and Equation~\ref{eq:sigma_eliashberg} by a simultaneous re-scaling of $\Sigma$ and $\Phi$ (see e.g., \cite{msv,Chubukov_2020a}). 
    \footnote{These singular terms  do affect superfluid stiffness $\rho_s$ at $T >0$ and extra care has to be exercised   to make sure that $\rho_s$ remain finite above a QCP (see e.g. Reference \cite{paper5}).}  
   Second,  $\Sigma (\omega_m)$ without the thermal piece vanishes in the normal state at first 
 Matsubara frequencies $\omega_m = \pm \pi T$  (the first Matsubara frequency rule, Reference \cite{Maslov_2012})
 has a  profound effect: $T_{c,0}$ does not vanish at $N = N_{cr}$ and 
 is actually non-zero even at the largest  $N$, where  $T_{c,0} = ({\bar g}/(2\pi)) (1/N)^{1/\gamma}$ (Reference~\cite{Wang2016}). The difference between $N > N_{cr}$ and $N < N_{cr}$ shows up in the behavior of the gap amplitude 
  below $T_{c,0}$: at $N \ll N_{cr}$ it monotonically increases with decreasing $T$, at $N \leq N_{cr}$ it displays a non-monotonic behavior but still remains finite at $T=0$, and at  $N > N_{cr}$  it vanishes at both $T = T_{c,0}$ and $T=0$ (see References \cite{paper2,Chubukov2020aa} for detail).

\begin{figure}[h]
\centering
\includegraphics[width=\textwidth]{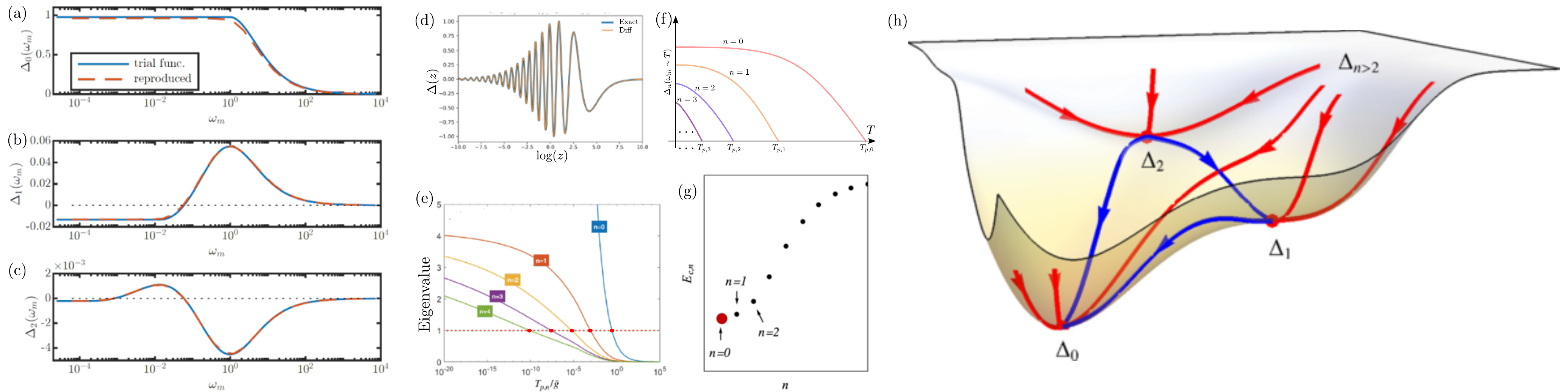}
\includegraphics[width=\textwidth]{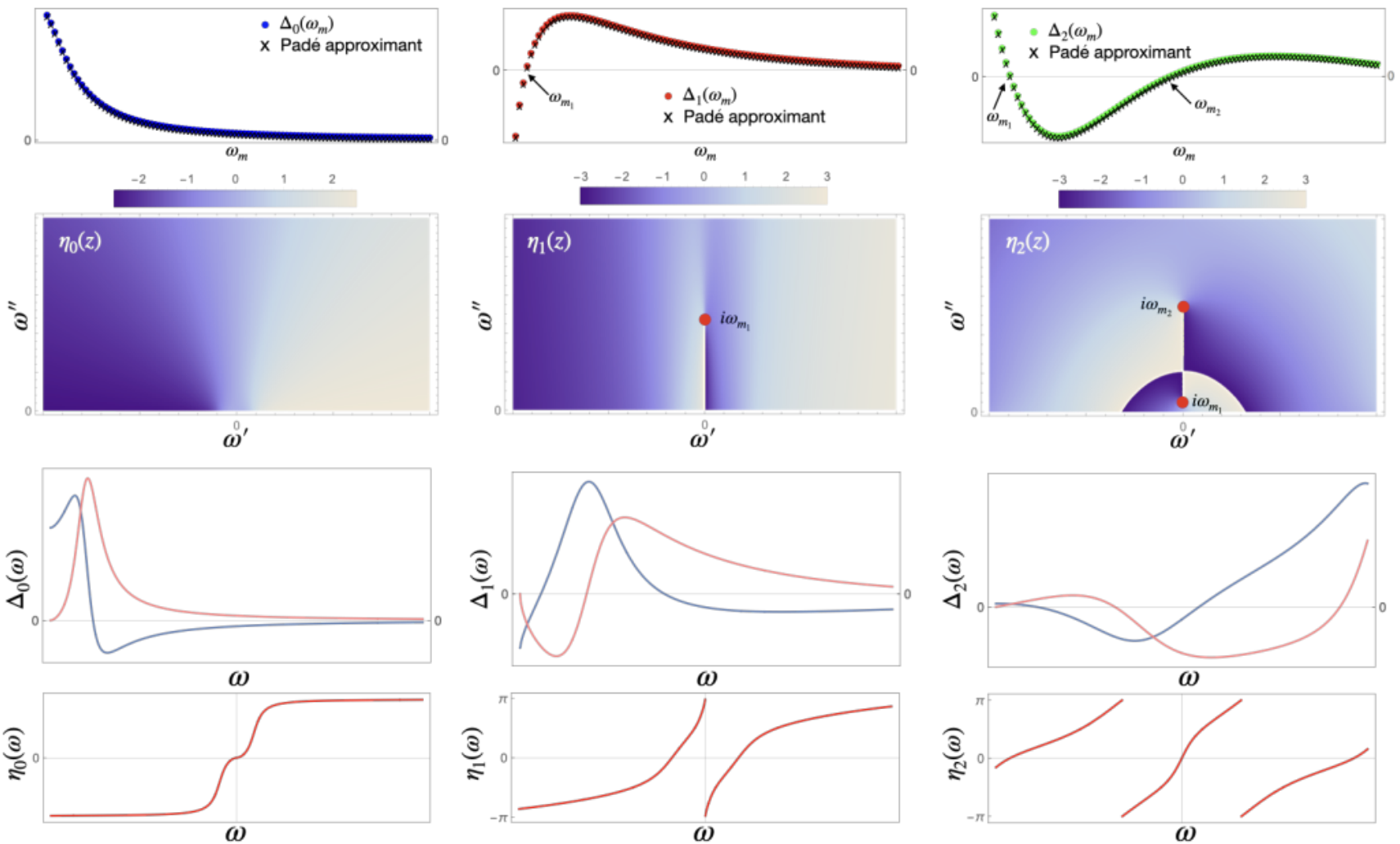}
\caption{ Upper panel: 
a) The solutions of the non-linear gap equation with $n =0,1$ and 2 sign changes. 
b) the solution with 
 $n=\infty$. The gap magnitude is  infinitesimally small and the number of oscillations is infinite, 
c) critical $T_{c,n}$ for different solutions for $N=1$,
d) the comparison of the exact solution
   of the integral equation
   and the solution of the approximate differential equation for $\Delta (z)$ ($z = (1-\gamma) \omega^{\gamma}_m$) ,
e) transition temperatures for the tower of solutions,
f) gap functions $\Delta_n (0)$ for different solutions vs $T$, 
g) condensation energies for the tower of solutions,
h)  free energy configuration in the functional space.  A solution 
   $\Delta_n(\omega_m)$ with $n >0$ is a saddle point of the free energy with $n$ unstable principle axes connected with the $n'<n$ solutions (the flows in the plot). The $n=0$ solution is a stable minimum. 
Lower panel: The vortex structure of the gap function $\Delta_n (\omega_m)$ and phase slips on the real axis. The continuation from Matsubara to real axis has been made using Pad\'e approximants. Panels from top to bottom:  - (i) $\Delta_n (\omega_m)$; (ii)  the phase of the gap function in the upper frequency half-plane  $\eta_n(z)$ = Im$[\log{\Delta (z)}]$, $z = \omega' + i \omega''$ (the locations of the  vortices are marked by red dots); (iii) real and imaginary parts of $\Delta (\omega)$ along the real axis; (iv)  variation of the phase of 
 $\Delta_n (\omega) = |\Delta_n (\omega)| e^{i\eta_n (\omega)}$ 
    along the real axis showing $n$ phase slips due to vortices.}
\label{fig:5}
\end{figure}

The existence of a finite superconducting $T_{c,0}$ at $N=1$  has been rigorously confirmed using the methods of  mathematical physics, which allowed to establish robust analytical upper and lower bounds on the exact transition temperature in the $\gamma$-model~\cite{Kiessling2025, Elezaby2025}. 

\subsection{Tower of solutions $\Delta_n (\omega_m)$} 
\label{sec:tower}

We now return to the issue of the number of solutions of the non-linear equation for $\Phi (\omega_m)$ (and $\Delta (\omega_m)$) at $T=0$. 
We start with a heuristic argument for the existence of multiple/infinite series solutions.  For this we 
 convert the  linearized gap equation 
 for $\Phi (\omega_{m})$ at $T=0$, Equation~\ref{eq:phi_eliashberg_l_1}, into differential equation by replacing  
  $|\omega_m - \omega^{'}_m|^\gamma$ by ${\text {max}} (|\omega_m|,|\omega^{'}_m|)$, like we  did to derive Equation~\ref{eq:24}.   Introducing $z = (1-\gamma)  \omega^\gamma_m$, we  express  Equation~\ref{eq:phi_eliashberg_l_1}  for $\omega_m >0$ as 
\begin{equation}\label{eq:8z}
 \Phi (z)=\left(\frac{1}{4}-b^{2} \right)\left[\frac{1}{z}\int_{0}^{z} \frac{\Phi (x)dx}{1+x}+\int_{z}^{\infty }\frac{\Phi (x)dx}{x(1+x)}  \right],
\end{equation}
where 
\begin{equation}
  b = \frac{1}{2} \sqrt{\frac{N-N_{cr}}{N}}, ~~ N_{cr} = 4(1-\gamma)/\gamma
 \label{eq:17}
\end{equation}
This $N_{cr}$ coincides with Equation~\ref{eq:N_cr} at small $\gamma $.
For $N>N_{cr}$, $b$ is real and $b\in [0,1/2)$, while for $0<N<N_{cr}$, $b=i\tilde{b}$ is imaginary and $\tilde{b}\in [0,\infty )$. 
 Notice  that Equation~\ref{eq:17} 
 is symmetric under $b\rightarrow -b$.
Differentiating Equation~\ref{eq:8z} twice over $z$, we obtain second order differential equation for $\Phi (z)$ in the form
\begin{equation}
\frac{d}{dz} \left[z^2 \frac{d \Phi(z)}{dz}\right] + \left(\frac{1}{4}-b^{2} \right) \frac{\Phi (z)}{1+z} = 0
\label{eq:9}
\end{equation}
Under the change of variables $z=1/y$,  this differential equation becomes a Schr\"odinger-like equation
\begin{equation}
\frac{d^2 \Phi (y)}{dy^2}  + \frac{\frac{1}{4}-b^{2} }{y(1+y)}  \Phi (y) = 0
\label{eq:9_1}
\end{equation}
  However, this equation
      has to be treated with caution as it was obtained by differentiation of the original  Equation~\ref{eq:8z}. 
      This procedure is prune to introduce parasitic  solutions. 
      In order to avoid them,  one has to define the space of functions to which the true solutions must belong. 
       For our case,  the   requirement is that $\int_0^\infty  dz \Phi^2 (z)/(1+z)$ must be convergent (Reference \cite{Abanov2020}). 
           This requirement establishes the norm on the space of functions,           completing the mapping of Equation~\ref{eq:9} to the Schr\"odinger-like problem.

 The two linearly independent solutions of Equation~\ref{eq:9} are~\cite{Abanov2020}
\begin{equation} \label{eq:Phi1Phi2}
\Phi_{\pm } (z) = H_{\pm b} (z),
\end{equation}
where
\begin{equation}
 H_b (z) = \frac{1+z}{z^{1/2-b}} \frac{\Gamma\left(\frac{1}{2} +b\right)\Gamma\left(\frac{3}{2} +b\right)}{\Gamma\left(1 +2b\right)} {_2}F_{1} \left[\frac{1}{2} +b, \frac{3}{2} +b, 1 +2b, -z\right]
 \label{eq:14_1}
\end{equation}
  and ${_2}F{_1} [...]$ is a Hypergeometric function.  At small $z$,   $H_b (z) \sim  1/z^{1/2-b}$. At large $z$, $H_b (z) = 1+\left(\frac{1}{4}-b^{2} \right) \log{z}/z+O(1/z)$. One can easily check, that the solutions with a real $b$ are not normalizable 
  (the norm has a power-law divergence at $z \to 0$ for one of the solutions and logarithmic divergence at   $z \to \infty$ for both).  As a result, there is no solution of the homogeneous linearized gap equation at $N > N_{cr}$.    At $N < N_{cr}$, when  $b=i\tilde{b}$ is imaginary, the situation is different.  
   Now each solution converges at $z=0$, and a convergence at  $z = \infty$ is achieved by  taking the  linear combination $\Phi (z)=\Phi_{-}(z)-\Phi_{+}(z)$ as the solution (for a more accurate treatment of the limit $z \to 0$ see Reference \cite{Abanov2024}).  As a result,  the linearized gap equation  does have a solution that satisfies the normalizability condition.   
      In Reference \cite{Abanov2020} we went further and obtained the exact solution of the original integral equation for $\Phi (z)$,  which also satisfies the normalizability condition.  We show  this solution along with the solution of the differential equation, Eq.  (\ref {eq:Phi1Phi2})  for $\gamma =0.01$ in  \textbf{Figure \ref{fig:5}d}. We see that they essentially coincide. We verified that the two solutions remain close  for all $\gamma <1$.  We emphasize that the solution of the linearized gap equation exists not only at $N = N_{cr} -0$, but at all  $N \leq N_{cr}$. This is another feature qualitatively different from BCS/E theory. 
      
   Consider next the non-linear gap equation.   One can easily  make sure that  its solution
   $\Phi (\omega_m)$  tends to a finite value $\Phi (0)$ at $ \omega_m  \to 0$. .  A natural way to rationalize such   solution is to 
    take the oscillating solution of the linearized gap equation, cut it at $ \omega^*_m \sim \Phi (0)$ and set the solution to be $\Phi (0)$ at smaller $\omega_n$. This can be done more rigorously by identifying $\omega^*_m$   with a frequency at which the oscillating solution of the linear gap equation has zero frequency derivative.   However, because of oscillations, there is an infinite number of $\omega^*_m$ at which the solution of the linearized gap equation has zero derivative.  
     By our logic,  there must be  then an infinite number of the solutions $\Phi_n (\omega_m)$ (and corresponding $\Delta_n (\omega_m)$). The solution with $n=0$ has zero sign changes, the solution with $n=1$ has one sign change at some $\omega_m$ along the upper half of the Matsubara frequency, the solution with $n=2$ has two sign changes, and so on. The magnitude of the gap decreases with increasing $n$. The limit $n= \infty$ is the solution with infinitesimally small $\Phi_n (\omega_m)$ and infinitely many sign changes. This is precisely the solution of the linearized gap equation, and this explains why the solution of the linearized gap equation (understood as the limiting solution of the non-linear gap equation)  exits for all $N \leq N_{cr}$.       
    
In \textbf{Figure \ref{fig:5}a-c, e-f} we show the results of the numerical solution of the non-linear gap equation both at  $T=0$ and at finite $T$ (References \cite{paper2,Chubukov2020aa}). 
 These results show that there indeed exists a tower of solutions $\Phi_n (\omega_m)$ ($\Delta_n (\omega_m)$) , each with  its own $T_{c,n}$. 
   The free energies of different solutions are different. For the pure $\gamma$-model,  the absolute minimum  is for  the $n=0$ solution,  \textbf{Figure \ref{fig:5},g}.
     other  solutions are saddle points of the  energy functional,  \textbf{Figure \ref{fig:5}h}. However, in the presence of extra interactions, the minimum may shift to $n >0$ (see Reference \cite{yu2026}). 

The solutions with different $n$  are topologically distinct. 
 A gap  function  $\Delta_{n}(\omega )$ has $n$ zeros on the positive Matsubara axis.  The corresponding 
    retarded $\Delta^{ret}_{n}(z)$ 
 is an analytical 
function  in the upper half-plane of frequency. The  number of zeros of such a function, $Z_n$,  is a topological invariant. It can  be expressed  as 
\begin{equation}\label{eq:numberOfZeros}
 Z_n=\frac{1}{2\pi i}\int_{-\infty }^{\infty }\partial_{\omega }\log (\Delta^{ret}_{n}(\omega ))d\omega 
\end{equation}
  and is the winding number of the phase of $\Delta^{ret}_{n}(\omega )$. We illustrate this in the lower panel in \textbf{Figure \ref{fig:5}}.
  
We re-iterate that the  existence of a tower of topologically distinct solutions is the ultimate consequence of 
   singular dynamical interaction with the lower cutoff $\Lambda =0$.  When $\Lambda$ becomes finite and increases, the 
     solutions with $n >0$ disappear one by one, starting from the solutions with the largest $n$ (Reference \cite{paper2}).

\section{CONCLUSIONS}

In this review, we discussed  the interplay between  non-Fermi liquid and superconductivity in a system with a singular dynamical interaction. Examples are quantum dots described by Yukawa-SYK model, itinerant metals near a QCP towards spin or charge  order,  and metallic systems not near a QCP in which coupling is strong enough to make the effective interaction dynamical and singular, but not too strong to localize itinerant carriers.   

The would be normal state for such systems is non-FL down to an energy $\Lambda$ which we set to be far smaller than the effective interaction ${\bar g}$ that gives rise to non-FL and to pairing.  We showed that in this situation the separation of scales,  present in BCS/E theories, breaks down and Cooper logarithm becomes inefficient.  Without Cooper logarithm, there is no superiority of the particle-particle channel over the particle-hole one, and both should be treated on equal footings.  In physical terms, this requires treating on equal footing the tendency towards 
 superconductivity and towards non-FL.     
 
We showed that  at $\Lambda \to 0$, ${\bar g}$ is the only parameter with dimension of energy and in units of ${\bar g}$  the competition between non-FL and superconductivity is a universal problem, whose outcome depends on the two numbers: relative strength of the interaction in the particle-particle and particle-hole channels, which we labeled as $1/N$, and the exponent $\gamma$ in the  power-law frequency dependence of both interactions.

We  showed that at a given $\gamma$, superconductivity develops only when $N$ becomes smaller than the threshold value $N_{cr}$. At larger $N$,  the  ground state remains a non-FL.   We also showed that in the absence of Cooper logarithm,
 the origin of superconductivity   qualitatively different from that in BCS/E theory and is associated with the appearance of a complex exponent for the pairing susceptibility. 
  In another qualitative distinction from BCS/E theory, we showed  that the gap equation at $T=0$ has an infinite set of   topologically distinct solutions. These solution disappear one by one once the pairing interaction becomes non-singular (massive). 

Overall, these results imply that superconductivity in a system with a singular dynamical interaction is not 
  merely a quantitative modification of the BCS/E theory, but a radically new narrative 
   for quantum matter, governed by its own universal rules.

\begin{summary}[SUMMARY POINTS]
\begin{enumerate}
\item  In systems with {\em singular dynamical} interaction down to a small cutoff,  the hierarchy of the energy scales changes and the coupling ${\bar g}$  becomes the only relevant energy scale. 
\item Cooper logarithm, which plays crucial role in BCS and Eliashberg theories, becomes ineffective and superconductivity competes on equal footings with the tendency towards non-FL ground state.
\item The competition is a universal phenomenon  and the outcome is determined by two numbers: 
relative strength of the interaction in the particle-particle and particle-hole channels, which we labeled as $1/N$, and the exponent $\gamma$ in the  power-law frequency dependence of both interactions.
\item  For any $\gamma$,  is a threshold value of $N$ separating superconducting and non-FL ground states. 
\item The pairing gap $\Delta (\omega )$ strongly depends on $\omega $ already at frequencies comparable to
$\Delta (0)$.
\item There exists an infinite number of topologically distinct solutions of the non-linear  gap equation, $\Delta_n (\omega)$,  $n=0,1\dots$; each has its own $T_{c,n}$.  
   For a pure $\gamma$-model,  topologically trivial  $\Delta_{0}(\omega )$ is a minimum of the energy functional. Other solutions are 
   saddle points.
\end{enumerate}
\end{summary}

\begin{issues}[FUTURE ISSUES]
\begin{enumerate}
\item \textbf{Superfluid Density and role of phase fluctuations:}  The analytical results rely on the Eliashberg approximation. A critical open question is how to extend theoretical description into the regime where vertex corrections are large and, as the consequence, phase fluctuations are strong. This is certainly the case 
      for models with $\gamma >1$ unless one takes the double limit in which $E_F$ tends to infinity  simultaneously with  $\Lambda \to 0$.   Future work must systematically incorporate vertex corrections and the full renormalization of the boson propagator to determine how to extend the theory beyond the Eliashberg approximation.

\end{enumerate}
\end{issues}

\section*{DISCLOSURE STATEMENT}
The authors are not aware of any affiliations, memberships, funding, or financial holdings that might be perceived as affecting the objectivity of this review.

\section*{ACKNOWLEDGMENTS}
 We thank  B. Altshuler,  R. Combescot, K. Efetov, R. Fernandes,  A. Finkelstein, E. Fradkin, A. Georges, S. Hartnol, S. Karchu, S. Kivelson, I. Klebanov,  S-S. Lee, G. Lonzarich, D. Maslov, F. Marsiglio, M. Metlitski, W. Metzner, A. Millis, D. Mozyrsky, C. Pepin, V. Pokrovsky,  N. Prokofiev,  S. Raghu,  S. Sachdev,  T. Senthil, D. Scalapino, Y. Schattner, D. Son, G. Tarnopolsky, A-M Tremblay, A. Tsvelik,  G. Torroba,  E. Yuzbashyan, and J. Zaanen for useful discussions.  We are particularly thankful to our collaborators in research on non-BCS superconductivity  E. Berg, D. Chowdhury,  L. Classen,   A. Klein, J. Schmalian, Y. Wang, Y. Wu and S-S Zhang. We thank S-S. Zhang for making the plot for Fig. 2a.  The work by  AVC was supported by the NSF DMR-2325357. 
 The work by ArA was supported by the DOE-Office Of Science, DE-SC0026038.

\bibliographystyle{ar-style4} 
\bibliography{nonBCSbib.bib,nonBCSbib2.bib}
\end{document}